\documentclass[aip,pop,psfig,twocolumn,showpacs,superscriptaddress,nobibnotes,
reprint]{revtex4-1}
\usepackage{color}
\usepackage{graphics}
\usepackage{epsfig}
\usepackage[normalem]{ulem}
\usepackage{cancel}
\usepackage{float}

\usepackage[outercaption]{sidecap}

\begin{document}
\title{Spiral Waves in Driven Dusty Plasma Medium: Generalized Hydrodynamic Fluid Description}

\affiliation{Institute for Plasma Research, HBNI, Bhat, Gandhinagar - 382428, India}
\author{Sandeep Kumar} 
\email{sandeep.kumar@ipr.res.in}
\author{Bhavesh Patel}
\author{Amita Das} 

\begin{abstract} 
\paragraph*{}
Spiral waves are observed in many natural phenomena. They have been extensively represented by the mathematical FitzHugh-Nagumo (FHN) model [Barkley et al., Phys. Rev. A $\bf{42}$, 2489 (1990)] of excitable media. In incompressible fluid simulations also excitation of thermal spiral waves have been reported by Li et al. [Phys. of Fluids $\bf{22}$, 011701 (2010)]. In the present manuscript the spatiotemporal development of spiral waves in the context of weak and strong coupling limits have been shown. While the weakly coupled medium has been represented by a simple fluid description, for the strong coupling a generalized visco - elastic fluid description has been employed. 
The medium has been driven by an external force in the form of a rotating electric field. It is shown that when the amplitude of force is small the density perturbations in the medium are also small. In this case the excitations do not develop as a spiral wave. Only when the amplitude of force is high so as to drive the  density perturbations to nonlinear amplitudes does the spiral density wave formation occur. The role of forcing frequency, the effect of strong coupling and sound velocity of medium on the formation and evolution of spiral waves have been investigated in detail. 

\end{abstract} 
\maketitle 
\section{Introduction}
\label{intro}
Rotating spiral waves are ubiquitous structures found in a wide range of natural and laboratory scenario. For instance, Belousov-Zabotinsky (BZ) reaction\cite{Keener}, excitable reaction-diffusion media\cite{Barkley, Bar}, liquid crystals\cite{Frish}, cardiac tissue\cite{Karma_tissue, Davidenko}, rotating fluids\cite{Li}, spiral galaxy\cite{Elmegreen, Springel}, coupled oscillators\cite{Martens,Shima}, etc.\cite{Kruse, Clar, Sun_2008, Huang}, all demonstrate the existence of spiral waves. The self organization of excitations in the form of spiral wave patterns continues to remain an intriguing topic. It has often been interpreted on the basis of an  interplay between propagator and controller fields in the excitable 
medium. The spiral wave tip can rigidly rotate or meander depending upon control parameters of the medium. A vast amount of literature is present in which people 
claim that meandering occurs via a Hopf bifurcation\cite{Barkley,Hakim}. Mathematically, FitzHugh-Nagumo (FHN) model has been widely employed for the  spatiotemporal development of spiral waves in excitable media\cite{FitzHugh, Nagumo, Barkley_1990, Pertsov}. For incompressible fluid system the thermal spiral wave pattern has been observed resulting from temperature gradient excitations \cite{Li}. In the present manuscript we show that a compressible fluid system can also be forced to form spiral wave density patterns. These waves are shown to propagate in a spiral pattern even after the forcing is switched off.  
 
 For definiteness we consider dusty plasma medium for our study. A dusty plasma is mixture of highly charged (mostly negative) and heavy ($10^{13} - 10^{14}$ times heavier than the ions) dust grains along with electron and ion species. A typical dust particle of micron size has approximately -$10^{4}$e charges. 
At slow dust time scales the inertialess response of electrons and ions are considered which essentially follow a Boltzmann distribution. The inclusion of heavy dust species makes such a plasma exhibit a rich class of collective phenomena. Depending on the value of its coupling parameter it can have both fluid like viscous as well as solid like elastic traits wherein it preserves memory of its past configurations for some amount of time. This has led to the adoption of visco - elastic fluid depiction in terms of Generalized Hydrodynamic (GHD) fluid model\cite{Kaw_1998, Kaw_2001}. 
The dusty plasma medium has been shown to exhibit a variety of normal modes such as the longitudinal acoustic\cite{Rao, Thompson, Thomas, Pieper, Rosenberg, Barkan} and transverse shear waves\cite{Pramanik, das_2014, Dharodi, Nunomura}. In the nonlinear regime the dusty plasma medium can excite self-sustained non-linear propagating waves that can form solitons\cite{Sandeep_KDV, Sanat_soliton}, shocks \cite{Samsonov, Sumita, Surabhi}, and vortices\cite{Bharuthram, Piel_2002, Shukla_2003, Hasegawa, Schwabe_2012} etc. Experimental studies by externally driving the medium by energetic particles and or external rotating electric fields (REF) have also been considered\cite{Hollmann, Danielson, Nosenko, Worner}. In this work we numerically study the response of the dusty plasma medium in the presence of external forcing by a rotating electric field using the generalized hydrodynamic model. Both weak (wherein the equations reduce to simple fluid description) and strong coupling regimes have been investigated in detail. 

The paper has been organized as follows. Section II gives a brief description of the Generalized Hydrodynamic model and its numerical implementation. In section III we report the observation of the excitation of spiral density waves in the presence of forcing. The salient features of spiral density wave with respect to various parameters of the medium and forcing characteristics are discussed in detail in various subsections. Section IV provides the summary and conclusion on the study. 

\section{Governing Equations  } 
\label{ghd_fluid}
It is well known that the dusty plasma medium behaves like an ordinary fluid when the coulomb coupling parameter $\Gamma$ is of the order of unity. For $\Gamma > \Gamma_c$, where $\Gamma_c \sim170$, it condenses in crystalline state. 
For intermediate values of $\Gamma$ it behaves like a visco - elastic fluid which has often been depicted by the GHD model description.

The equations depicting the evolution of the visco - elastic dusty plasma medium in the presence of external forcing is represented by the coupled set of equations of GHD and continuity equation for velocity and density evolution \cite{Sanat_KH}. 
\begin{eqnarray}
& & \left[1 + \tau_ m (\frac{\partial}{\partial t} + \vec{v}\cdot \nabla )\right] \times  \nonumber \\
& & \left[\left(\frac{\partial \vec v}{\partial t} + \vec v \cdot \nabla \vec v \right) + \frac{\nabla P}{n_d} -\nabla \phi- \text{F}_{\text{rot}}\right] = \eta \nabla^2 \vec v
\label{ghd}
\end{eqnarray}
\begin{equation}
   {\frac{\partial n_d}{\partial t} + \nabla \cdot (n_d \vec v) = 0  }
   \label{cont}
   \nonumber
   \end{equation}
Where we have chosen the external force to be operative in a central circular patch with radii $r_0$ of the simulation box. Thus  
\begin{eqnarray}
\text{F}_{\text{rot}} &=& \text{A}\cos(\omega_f t)\hat{x} + \text{A}\sin(\omega_f t)\hat{y};  \hspace{0.2 in} {\text{r} <\text{r}_0} \nonumber \\
\text{F}_{\text{rot}} &=& 0; \hspace{0.2in} \text{otherwise} 
\label{forcing}
\end{eqnarray}
Where, A and $\omega_f = {2\pi}/{T_f}$ are amplitude and angular frequency of force, respectively. The magnitude of force is constant but direction rotating in time. Also $P$ here is the dust pressure (for which equation of state is used) and $\phi$ represents the scalar potential. The scalar potential is determined by the Poisson's equation: 
\begin{equation}
   {\nabla^2 \phi = - 4\pi e(n_d + \mu_e \exp(\sigma_e \phi) - \mu_i \exp(-\phi))}
   \label{poiss}
   \end{equation}
with parameters $\sigma_e = T_i/T_e $, $\mu_e = {n_{e0}}/{Z_d n_{d0}}$ and $\mu_i = {n_{i0}}/{Z_d n_{d0}}$, where $T_i$ and $T_e$ are the ion and electron temperature, $n_{i0}$, $n_{e0}$ and $n_{d0}$ are the equilibrium density of ion, electron and dust fluid, respectively and $Z_d$ is the negative charge on each dust particle. In unperturbed equilibrium situation dusty plasma medium satisfy the quasineutrality condition $n_{i0} = Z_d n_{d0}+ n_{e0}$. Here, we have considered the Boltzmann distribution for electrons and ions on dust response time scale so as to have:   
\begin{equation}
{n_e = \mu_e \exp (\sigma_i \phi)}; 
\hspace{0.5cm} {n_i = \mu_i \exp(-\phi)}, 
\end{equation}
It should be noted that the GHD model Eq. (\ref{ghd}) reduces to the momentum equation for a viscous compressible fluid in the limit when $\tau_m \rightarrow 0$, for which $\eta$ represents the viscosity. A finite  $\tau_m$ represents the time for which the fluid retains memory of its past configurations arising due to elastic behavior resulting from strong coupling features. Thus if one is looking for a phenomena with time scales for which the condition $\tau_m \frac{d}{dt}$ $<< 1$, is satisfied the dust fluid would exhibit essentially a behavior of normal 
viscous fluid. However, at faster time scales for which $\tau_m \frac{d}{dt}$ $> 1$, the dust fluid retains its memory and characteristic new 
elastic response can be observed. The variables $v$, $\phi$ and $n_s$ ($s = e, i, d$) are the dust fluid velocity, potential, and number density of the charged species (electrons, ions, and dust), respectively. The normalized number densities are $\bar{n_d} = {n_d}/{n_{d0}}$, $\bar{n_i} = {n_i}/{n_{i0}}$, $\bar{n_e} = {n_e}/{n_{e0}}$. Time and length are normalized by $\omega_{pd}^{-1}$ and $\lambda_D$ $(=k_B T_i/4\pi Z_d n_{d0} e^2)^{\frac{1}{2}}$. The normalized scalar potential is $\bar{\phi} = Z_d e \phi/k_B T_i$. The pressure is determined using equation of state $P = \mu_d \gamma_d n_d k_B T_d$. Here $\mu_d = \frac{1}{T_d} \frac{\partial P}{\partial n_d}|_{T_d}$ is compressibility parameter and $\gamma_d$ is adiabatic index. The parameters $\mu_d$, $\tau_m$, and $\eta$ are supposed to be empirically related to each other\cite{Ichimaru_1987, Kaw_1998, Das_RT}. 
  
A flux corrected scheme proposed by Boris et al., \cite{Boris} was used to evolve Eqs. (\ref{ghd},\ref{cont}). Since the scheme numerically solves the continuity form of equations with source and sink terms, we split Eq. (\ref{ghd}) as two separate equations of the following form:
\begin{equation}
{\left[1 + \tau_ m (\frac{\partial}{\partial t} + \vec{v}\cdot \nabla )\right]\vec \psi = \eta \nabla^2 \vec v}
\nonumber 
\end{equation} 
\begin{equation}
{\left(\frac{\partial}{\partial t} + \vec{v}\cdot \nabla \right)\vec v + \alpha\frac{\nabla n_d}{n_d} - \nabla \phi -\text{F}_{\text{rot}} = \vec{\psi}}
\end{equation}
Where, $\alpha = \mu_d \gamma_d T_d/T_i Z_d$ represents the square of sound speed of the medium.

\section{Numerical Observations}
We carry out numerical simulation studies for the coupled set of Eqs. (\ref{ghd},\ref{cont}) along with Poisson equation (\ref{poiss}) in a 
2-D $x-y$ plane. The boundary condition are chosen to be periodic. The box dimension have been chosen as $L_x = 16$, $L_y = 16$. The dust fluid is chosen to have a zero velocity and homogeneous density distribution initially.  
A rotating electric field forcing is applied in a small circular domain at the center of $x-y$ plane with a radii $r_0 = 0.5$. 
Both cases, where forcing continues to be present throughout the simulation and/or is switched off after a certain duration has been considered in our 
studies. 

\subsection{Weak coupling: fluid regime}
\begin{figure}[h]   
 \includegraphics[height = 7.5cm,width = 9cm]{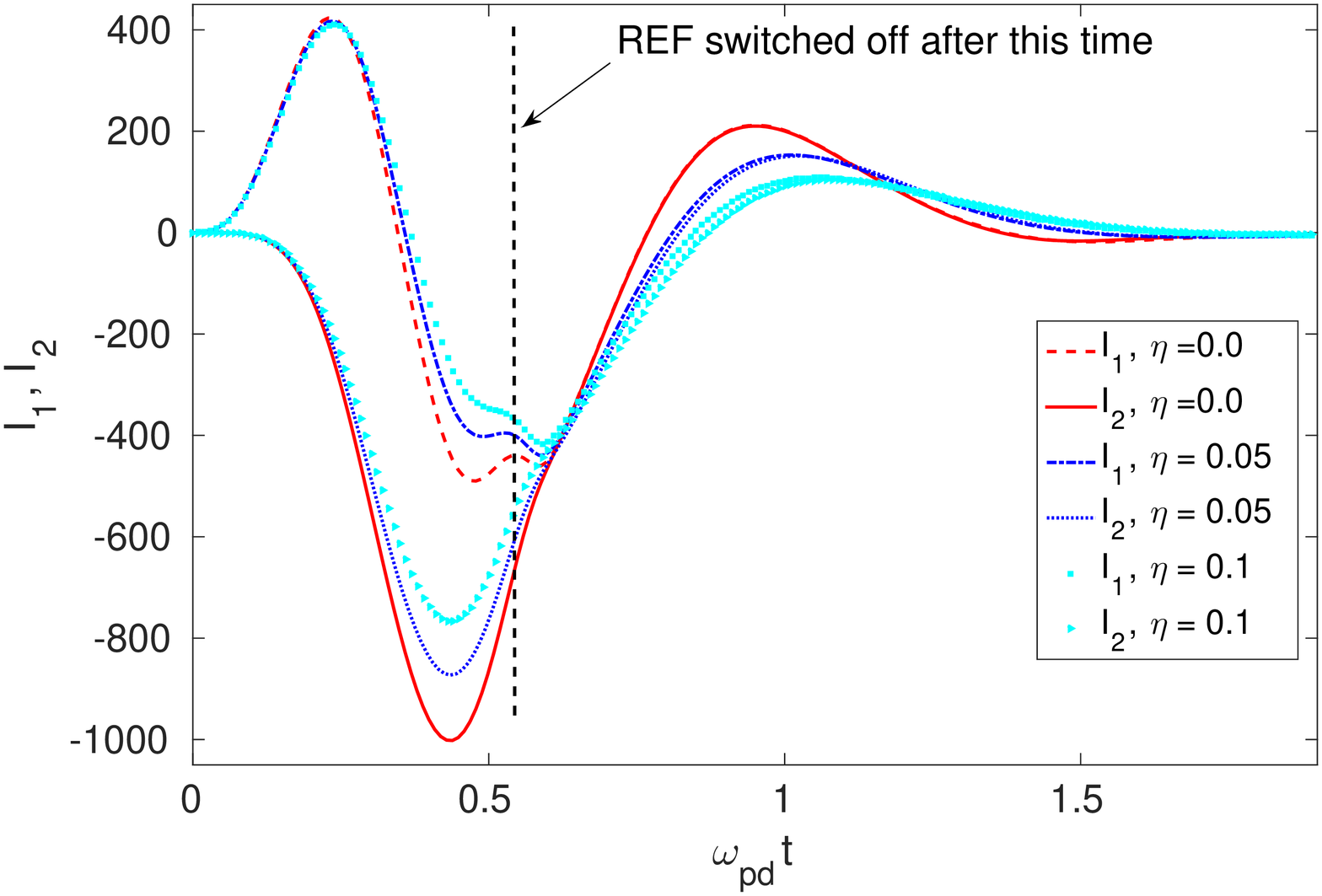}
                   \caption{ Time evolution of $I_1$ and $I_2$ (Eq. \ref{enstrophy-HD}) for compressible HD fluids with different values of $\eta$. Simulation parameters for this plot are $A = 10$, $\alpha = 5$ and $\omega_f = 10$. Here, applied force switched off after $\omega_{pd}t = 0.55$ time. In the plot $I_1$ and $I_2$ for different $\eta$ are represented by different line styles. There is perfect matching of $I_1$ and $I_2$ (dash and solid line, respectively) for the value of $\eta = 0$. }
        
                   \label{HD_inequality}	                       
        \end{figure}

In the weak coupling case the dust momentum equation satisfies the evolution equation of a normal charged fluid. The form of the equation can be recovered by choosing $\tau_m = 0$ in Eq. (\ref{ghd}). The evolution of vorticity in the compressible fluid regime (i.e. when $\tau_m =0$) can be obtained by taking the curl of Eq. (\ref{ghd}) along with the continuity equation Eq. (\ref{cont}). Using the the vector identity of $$ \vec{v} \cdot \nabla \vec{v} = \nabla(v^2/2) - \vec{v} \times (\nabla \times \vec{v})$$ and assuming that the pressure is simply a function of density through equation of state we have in 2-D: 
 \begin{equation}
 \frac{\partial }{\partial t} \left(\frac{\vec{\Omega}}{n_d} \right) + \vec{v}\cdot \nabla \left(\frac{\vec{\Omega}}{n_d} \right) = \frac{1}{n_d }\left(\nabla \times \vec{\text{F}}_{\text{rot}} + \eta \nabla^2 \vec{\Omega}\right)
 \label{vort-fluid-evo}
 \end{equation}
 Here, $\vec{\Omega} = \nabla \times \vec{v}$ is the vorticity. From Eq. (\ref{vort-fluid-evo}) in the absence of forcing and viscosity, the field $\vec{\Omega}/n_d$ is convected by the fluid. 
Integrating over space it can be shown that 
\begin{equation}
 \frac{d}{dt}\int  \left(\frac{\vec{\Omega}}{n_d} \right) dx dy = \int  \left(\frac{\vec{\Omega}}{n_d} \right) \nabla \cdot \vec{v}  \hspace{0.05in} dx dy
 \label{enstrophy-HD}
\end{equation}
We denote the left and right hand side of  Eq. (\ref{enstrophy-HD}) by $I_1$ and $I_2$, respectively. We have shown the evolution of $I_1$ and $I_2$ with time in Fig. \ref{HD_inequality}. Both $I_1$ and $I_2$ are initially zero as the medium is unperturbed in the beginning. The forcing is responsible for the generation of 
vorticity and density perturbations. The vertical line shows the time at which the forcing is stopped. It can be seen that for $\eta = 0$, once the forcing is stopped there is a close agreement between $I_1$ and $I_2$ as expected.
\begin{figure}[h]   
 \includegraphics[height = 7.5cm,width = 7.5cm]{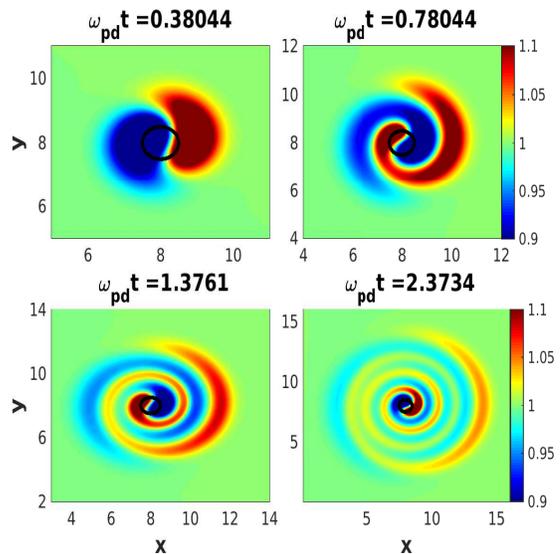}
                   \caption{Time evolution of the density for HD fluid. Here $\text{F}_{\text{rot}}$ applied for whole duration of simulation. For simulation, We have taken $A = 10$, $\omega_f = 10$, $\alpha = 5$ and $\eta = 0.1$. Small circle in the figure represent the region where REF applied. Density evolution of medium show the formation of spiral wave.}
        
                   \label{HD_time_evol_all_time}	                       
        \end{figure}        
 \begin{figure}[h]   
 \includegraphics[height = 7.5cm,width = 7.5cm]{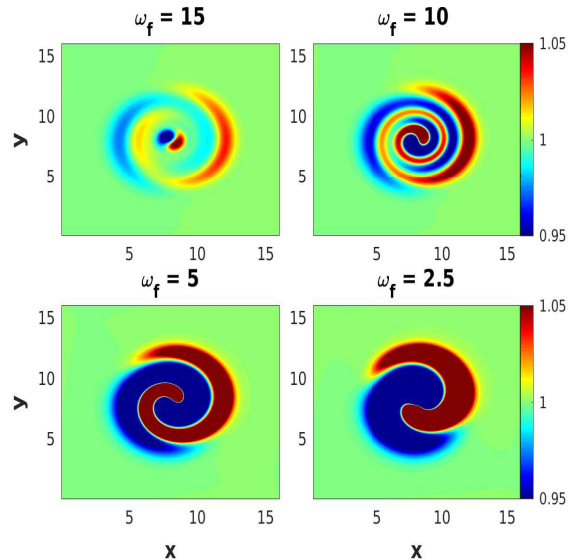}
                   \caption{ Spiral wave for different frequency of forcing. In this case rotating external forcing applied for whole duration of simulation. Simulation parameters for this plot are $A = 10$, $\eta = 0.1$, and $\alpha = 5$. Density for all subplots are taken at time $\omega_{pd}t = 1.60$.}
        
                   \label{HD_fluid_eff_freq_all_time}	                       
        \end{figure}
 \begin{figure}[h]   
 \includegraphics[height = 7.5cm,width = 7.5cm]{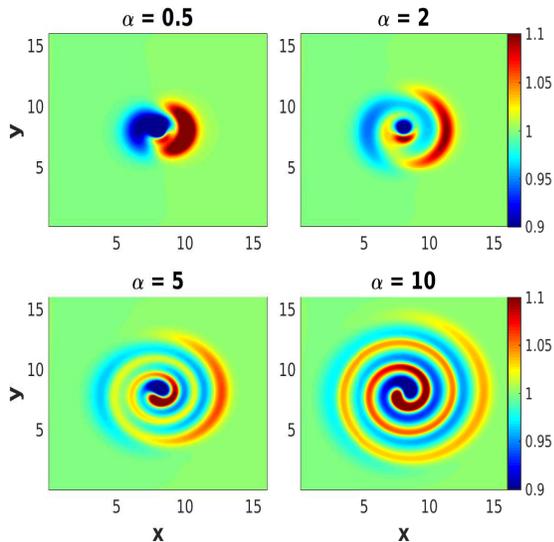}
                  \caption{ Dynamics of spiral wave for different value of sound velocity of medium ($\sqrt{\alpha}$). Simulation parameters for this plot are $A = 10$, $\omega_f = 10$, and $\eta = 0.1$. In this case rotating external forcing applied for whole duration of simulation. Density for all subplots are taken at time $\omega_{pd}t = 1.84$. Here by increasing $\alpha$ inner part of spiral wave breakup avoided. For higher value of $\alpha$, radial expansion is larger than azimuthal expansion.}
        
                   \label{HD_fluid_eff_alpha_all_time}	                       
        \end{figure}
It is, however interesting to observe the behavior of density perturbations in the 2-D plane, the snapshot of which has been shown at various time in Fig. \ref{HD_time_evol_all_time}. It shows a clear development of a spiral wave (density compression and rarefaction) with time. In this case the forcing is present throughout the simulation duration, however, it is applied only within a spatial region of $r< r_0 = 0.5$. The observed spiral wave, however, is extended beyond this spatial domain. Thus the spiral wave pattern are intrinsic response of the medium in the presence of such a forcing. The number of spiral rings at the various snapshots match with the number of forcing periods covered in that duration. For instance, the normalized forcing frequency $\omega_f = 10$ correspond to a time period of $T_f = 2\pi/\omega_f = 0.628$. Thus for the four subplots, we have $t/T_f = 0.605, 1.24, 2.91, 3.78$, which corresponds approximately to the number of rings that one observes in the subplots for this  figure. In Fig. \ref{HD_fluid_eff_freq_all_time} where we show the variation of the 2-D density plots with respect to the forcing frequency. From this figure too it is pretty evident from all the subplots except the first one that the spiral rings denote the number of forcing periods covered in a given duration for which the plot has been shown. In the first subplot of this figure the forcing frequency seems to be very high for the natural response of the 
 medium to keep pace. The natural response of the medium is typically the acoustic waves. It should be noted that the typical radial extent of the structure 
 for all the cases remains approximately the same as the acoustic speed in all the four cases of this figure is same. Since the rotation frequency is fast, the radial expansion is unable to keep pace with it and the spiral rings get smeared up to be distinguished clearly for the case of $\omega_f = 15$. For the other cases the number of rings are in agreement with the law mentioned above. 

\begin{figure}[h]   
 \includegraphics[height = 7.5cm,width = 7.5cm]{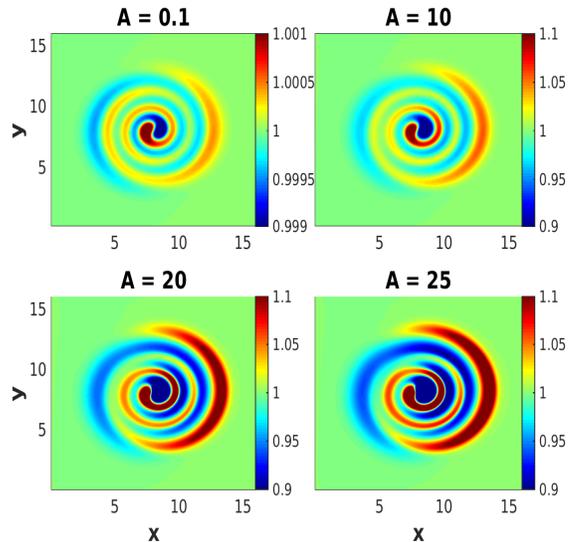}
                   \caption{Spiral wave with varying amplitude of forcing. In this case rotating external forcing applied for whole duration of simulation. Simulation parameters are $\omega_f = 10$, $\alpha = 5$ and $\eta = 0.1$. Density plots for all amplitudes are taken at time $\omega_{pd}t = 1.97$.}
        
                   \label{HD_eff_amp_all_time}	                       
        \end{figure}
\begin{figure}[h]   
 \includegraphics[height = 7.5cm,width = 7.5cm]{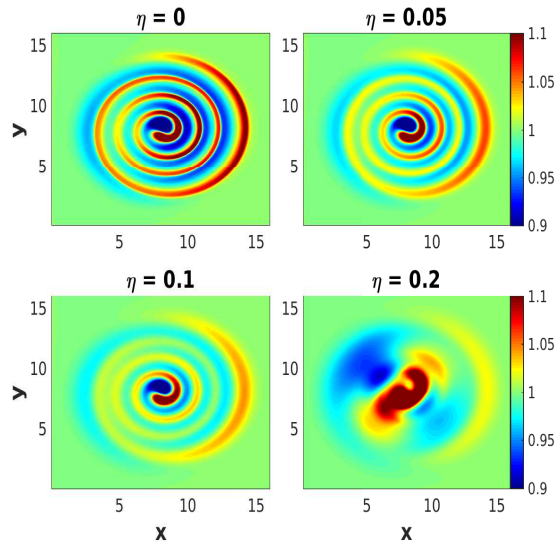}
                   \caption{Density plot of spiral wave for different values of $\eta$ in HD fluid. Simulation parameters in this plot are $A = 10$, $\alpha = 5$ and $\omega_f = 10$. Density plots for all values of $\eta$ are taken at time $\omega_{pd}t = 2.45$.}
        
                   \label{HD_dens_diff_eta}	                       
        \end{figure}
\begin{figure}[h]   
 \includegraphics[height = 7.5cm,width = 7.5cm]{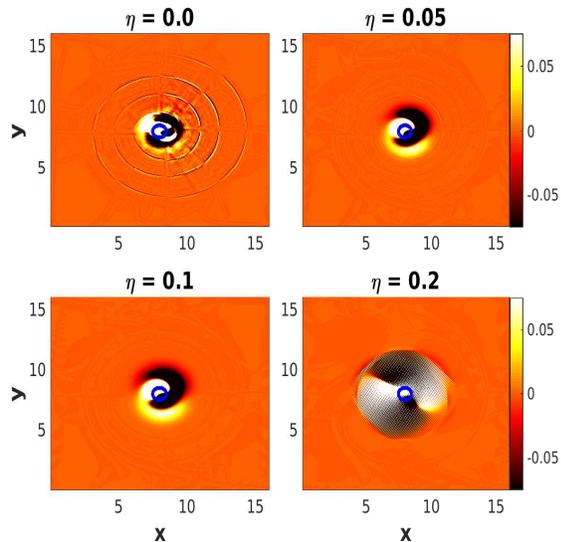}
                   \caption{$\Omega/n_d$ (vorticity) of the HD fluid medium for different values of $\eta$. Simulation parameters in this plot are $A = 10$, $\alpha = 5$ and $\omega_f = 10$. Vorticity plots for all values of $\eta$ are taken at time $\omega_{pd}t = 2.45$. From the figure it is clear that diffusion of source of vorticity increases with increase in the coefficient of viscosity of the medium.}
        
                   \label{HD_dens_diff_eta_vor}	                       
        \end{figure}
 It appears that the radial expansion of the structure is typically governed by the acoustic speed of the medium and the number of spiral rings by the forcing frequency. In fact for this case the value of $\alpha $ has been chosen as $\alpha= 5$. This typically corresponds to the acoustic speed (small corrections due to nonlinearity might exist at higher amplitudes) of $2.23$. This reasonable explains the radial expansion for $\omega_f = 10$ wherein the disturbance typically has propagated along $+\hat{x}$ direction from $r_0 = 8.5$ to $r = 12.4$ in a time duration of $t = 1.6$. The plot in Fig. \ref{HD_fluid_eff_alpha_all_time} shows variations with $\alpha$ also suggests that the radial expansion in our system is essentially governed by the value of $\alpha$. Another feature to note from Fig. \ref{HD_fluid_eff_freq_all_time} and Fig. \ref{HD_fluid_eff_alpha_all_time} is that for a proper unbroken spiral to form an appropriate combination of $\alpha$ and forcing frequency $\omega_f$ is required. This is because the radial expansion has to keep pace with the rotation. We have also observed the behavior of the spiral with respect to the amplitude $A$ of forcing. This has been shown in the plot of Fig. \ref{HD_eff_amp_all_time}. With increasing amplitude the density perturbations are stronger as the amplitude of density perturbations also increase. On the other hand in the presence of viscosity the density perturbations and $\vec{\Omega}/n_d$ die away as is shown in Figs. (\ref{HD_dens_diff_eta},\ref{HD_dens_diff_eta_vor}) and the spiral waves get damped as expected. Fig. \ref{HD_dens_diff_eta_vor} also elucidate that the source of spiral vorticity (small circular forcing region) diffused with increase in the viscosity of the medium which is also in agreement with the Eq. (\ref{vort-fluid-evo}). 
\subsection{Strong coupling: GHD regime}
We now present the evolution of the complete set of GHD fluid equations for the dusty plasma medium. Again the initial configuration of homogeneous plasma density with zero velocity in 2-D $x-y$ plane is chosen. The dust fluid is subjected to time dependent forcing within a central circular spatial domain of the 2-D simulation box. 
%

\begin{figure}[h]   
 \includegraphics[height = 7.5cm,width = 7.5cm]{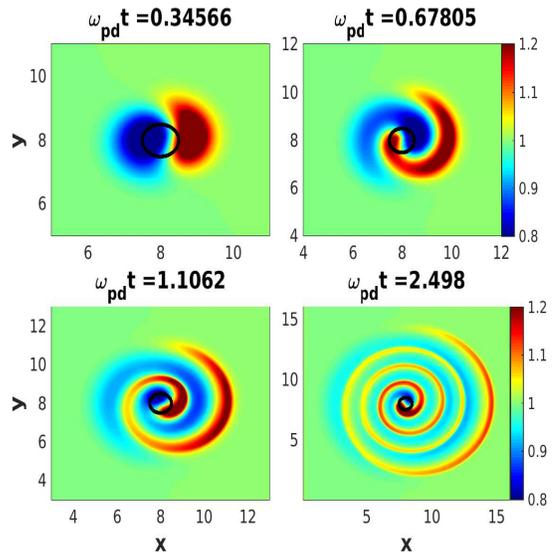}
                   \caption{Time evolution of the density for strongly coupled visco-elastic fluid. Here $\text{F}_{\text{rot}}$ applied for whole duration of simulation. For simulation, We have taken $A = 10$, $\omega_f = 10$, $\alpha = 5$,
                    $\eta = 5$ and $\tau_m = 20$. Small circle in the figure represent the region where REF applied.}
        
                   \label{GHD_time_evol_all_time}	                       
        \end{figure}

The snapshots of dust density evolution is shown at various times in Fig. {\ref{GHD_time_evol_all_time}}. For this case, the amplitude of the forcing $A =10$, the forcing frequency $\omega_f = 10$, $\eta = 5$,  $\tau_m = 20$ 
 and $\alpha = 5$ has been chosen. It should be noted that the periodicity of the forcing function being $10$ (in units of $\omega_{pd}$) the forcing function has completed several rotations at the time snapshot at which the four subplots of the Fig. {\ref{GHD_time_evol_all_time}} have been shown. The number of turns of the spiral rings in these snapshots are equal to the number of rotations of the forcing function here also similar to the weakly coupled case. However, even though the value of $\eta= 5$ is very high the spiral wave survives in this case when $\tau_m$ is finite. This is because $\eta$ in the presence of finite $\tau_m $ plays the role of elasticity of the medium. This is unlike the role of pure damping in the weakly coupled case. 
 \begin{figure}[H]   
 \includegraphics[height = 7.5cm,width = 7.5cm]{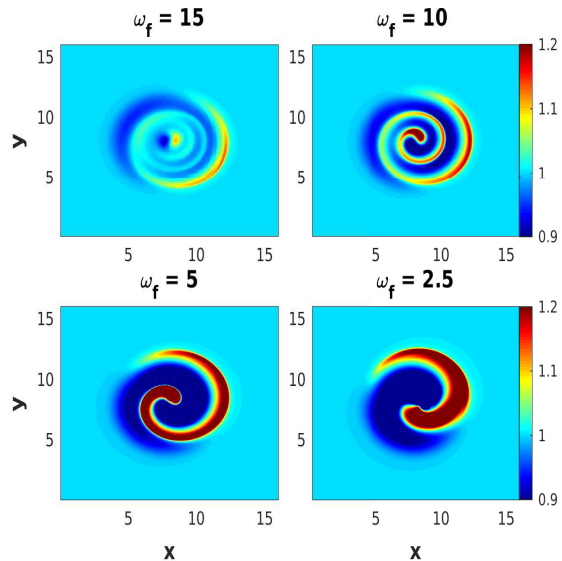}
                   \caption{ Spiral wave for different frequency of forcing. In this case too rotating external forcing applied for whole duration of simulation. Simulation parameters in this plot are $A = 10$, $\alpha = 5$, $\eta = 5$ and $\tau_m = 20$. Density for all subplots are taken at time $\omega_{pd}t = 1.55$.}
        
                   \label{GHD_eff_freq_all_time}	                       
        \end{figure}
 \begin{figure}[H]   
 \includegraphics[height = 7.5cm,width = 7.5cm]{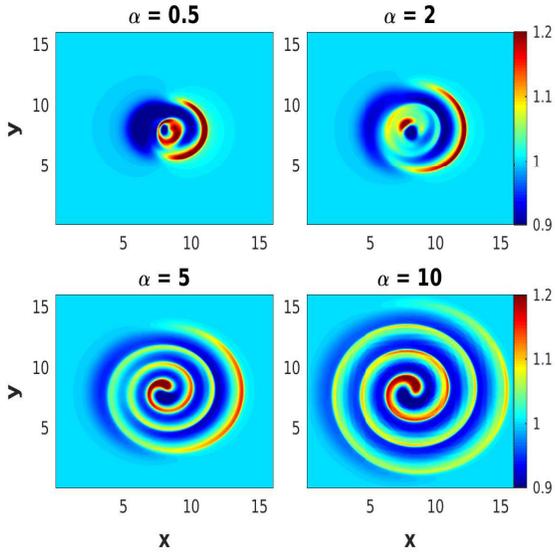}
                   \caption{ characteristic of spiral wave with varying sound velocity of the visco-elastic medium ($\sqrt{\alpha}$). Simulation parameters in this plot are $A = 10$, $\omega_f = 10$, $\eta = 5$ and $\tau_m = 20$. In this case rotating external forcing applied for whole duration of simulation. Density for all subplots are taken at time $\omega_{pd}t = 2.13$. Here by increasing $\alpha$ inner part of spiral wave breakup avoided. For higher value of $\alpha$, radial velocity is larger than angular velocity.}
        
                   \label{GHD_eff_alpha_all_time}	                       
        \end{figure}
\begin{figure}[h]   
 \includegraphics[height = 7.5cm,width = 7.5cm]{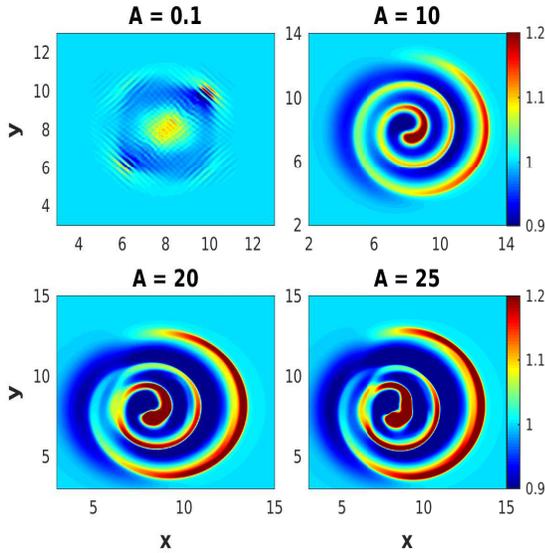}
                   \caption{Spiral wave behavior in GHD fluid with varying amplitude of forcing. In this case rotating external forcing applied for whole duration of simulation. Simulation parameters are $\omega_f = 10$, $\alpha = 5$, $\eta = 5$ and $\tau_m = 20$. Density plots for all amplitudes are taken at time $\omega_{pd}t = 1.76$.}
        
                   \label{GHD_eff_amp_all_time}	                       
        \end{figure}

\begin{figure}[h]   
 \includegraphics[height = 7.5cm,width = 7.5cm]{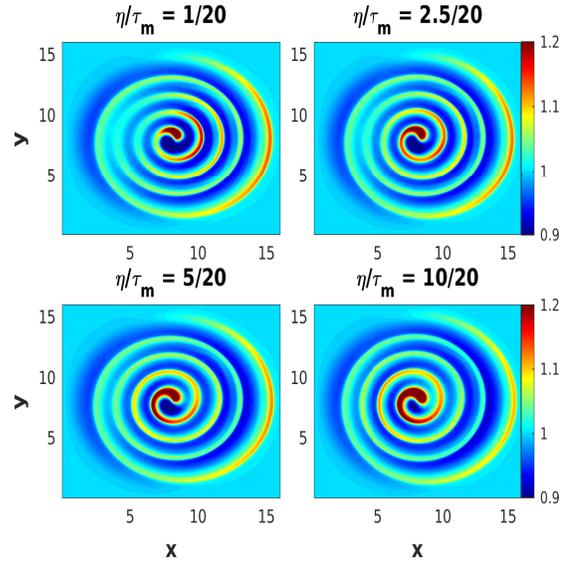}
                   \caption{Density plot of spiral wave with increasing strong coupling. Simulation parameters in this plot are $A = 10$, $\alpha = 5$ and $\omega_f = 10$. In this case too external REF applied for the whole duration of simulation. Density plots for all ratio of $\eta$ and $\tau_m$ are taken at time $\omega_{pd}t =  2.8068$.}
        
                   \label{GHD_dens_diff_eta_tau}	                       
        \end{figure}

\begin{figure}[h]   
 \includegraphics[height = 7.5cm,width = 7.5cm]{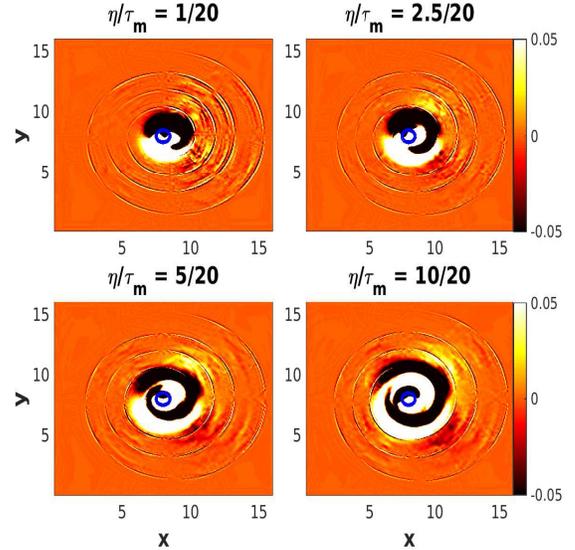}
                   \caption{$\Omega/n_d$ (vorticity) of the GHD fluid medium with increasing ratio of $\eta$ and $\tau_m$. Simulation parameters in this plot are $A = 10$, $\alpha = 5$ and $\omega_f = 10$. Vorticity plots for all ratio of $\eta$ and $\tau_m$ are taken at time $\omega_{pd}t = 2.8068$. From the figure it is clear that expansion of source of vorticity increases with increase in the ratio of $\eta$ and $\tau_m$.}
        
                   \label{GHD_vor_diff_eta_tau}	                       
        \end{figure}

\begin{figure}[h]   
 \includegraphics[height = 7.5cm,width = 7.5cm]{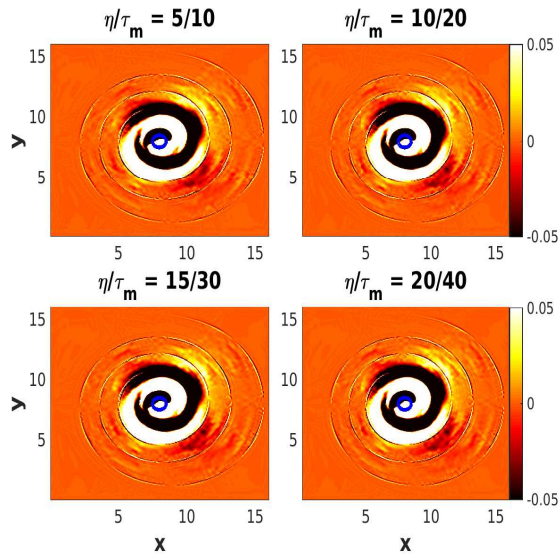}
                   \caption{$\Omega/n_d$ (vorticity) of the visco - elastic fluid for the same ratio of $\eta$ and $\tau_m$. Simulation parameters for this plot are $A = 10$, $\alpha = 5$ and $\omega_f = 10$. Vorticity plots for all ratio of $\eta$ and $\tau_m$ are taken at time $\omega_{pd}t = 2.8068$. Small circle in the figure represent the region where REF applied. From the figure it is clear that expansion of source of vorticity depends only upon the ratio of $\eta$ and $\tau_m$, not upon the individual values.}
        
                   \label{GHD_vor_same_eta_tau}	                       
        \end{figure}

%

  The behavior of the structure as a function of forcing frequency, $\alpha$ (representing the square of acoustic speed), amplitude ($A$) and strong coupling ($\eta/\tau_m$) are shown in Figs. \ref{GHD_eff_freq_all_time}, \ref{GHD_eff_alpha_all_time}, \ref{GHD_eff_amp_all_time} and (\ref{GHD_dens_diff_eta_tau}, \ref{GHD_vor_diff_eta_tau}, \ref{GHD_vor_same_eta_tau}), respectively. It is clear from Fig. \ref{GHD_eff_freq_all_time} that the number of rings gets decided by the forcing frequency. In this case too if the forcing frequency is very high the radial expansion of the medium is unable to keep pace with it and so in the $\omega_f =15 $ case the rings gets thinner to be distinguished clearly. When the frequency is low the radial width of the spiral arms are broad. This can be understood by realizing that the density perturbations in this case are forced at low frequency. The acoustic waves which can resonate at such low frequency would have longer wavelengths. It can be observed that good spirals are not clearly formed when the radial speed determined by $\alpha$ is small (Fig. \ref{GHD_eff_alpha_all_time} top subplots). A certain combination of $\alpha$ and forcing frequency determines a good spiral wave structure and its propagation with time. 

   At a low amplitude $A$ of forcing, one can observe that the spiral structure does not form as the perturbed density is too weak (Fig. \ref{GHD_eff_amp_all_time} for $A = 0.1$). When the value is increased to $A =10$, it can be observed that a good spiral wave structure gets formed. However, when the amplitude is increased further the density perturbation is high and the acoustic density perturbations would be nonlinear. This is visible from the bottom subplots of Fig. \ref{GHD_eff_amp_all_time}, where one can observe the formation of defects in the spiral structure.   
   
    We have also observed the behavior of the spiral wave with respect to strong coupling of the medium. This has been shown in the plot of Figs. \ref{GHD_dens_diff_eta_tau}, \ref{GHD_vor_diff_eta_tau} and \ref{GHD_vor_same_eta_tau}. From the Figs. \ref{GHD_vor_diff_eta_tau} and \ref{GHD_vor_same_eta_tau}, it is clear that with increasing strongly coupling (ratio of $\eta$ and $\tau_m$) the source of spiral vorticity (small circular forcing region) expanding.

\section{Summary and Conclusion} 
\label{result}
We have carried out the GHD visco - elastic fluid simulations for a driven dusty plasma medium. We study the response of the dust medium to an imposed rotating electric field 
in a small localized region and observed the formation of spiral waves. These spiral waves are observed to propagate radially outwards much beyond the spatial extent of the forcing. When forcing is weak and the density perturbations are in linear regime then we observe planar acoustic excitations. Only in nonlinear density perturbations does one observe spiral structure formation. Spiral wave have both angular as well as radial velocity. We have identified that the radial expansion velocity corresponds to the sound speed. The number of rings in the spiral correspond to the number of rotations of the forcing field at any given time. If the radial velocity is too fast then the rings are broad. However, when the forcing frequency is fast and the radial velocity is slow the spiral rings are sharp. When the radial velocity is too slow and is unable to keep pace with the forcing frequency the spiral structure suffers a break. For a proper unbroken spiral to form an appropriate combination of sound speed and forcing frequency is required. We have found that source of spiral vorticity expanding with increase in the strong coupling of the medium. We have observed only two armed (one of compression and other of rarefaction) spiral waves unlike Li et al.\cite{Li} multi-armed spiral waves. 
 
 Spiral wave formation are ubiquitously present in many natural phenomena in excitable media which requires a certain time duration to regain after a wave passes through it. In this case the dust density perturbations created by the forcing requires the response of the dust before it can be again perturbed by the forcing. The exciter and controller fields are believed to be important in the excitation of spiral waves. Here forcing induces vorticity as well density perturbations which together propagate like a spiral wave pattern.

\newpage
\bibliography{spiral_GHD_ref}
\end{document}